# Polar, Spherical and Orthogonal Space Subdivisions for an Algorithm Acceleration: O(1) Point-in-Polygon/Polyhedron Test

Vaclav Skala

*Abstract*—Acceleration of algorithms is becoming a crucial problem, if larger data sets are to be processed. Evaluation of algorithms is mostly done by using computational geometry approach and evaluation of computational complexity. However in today's engineering problems this approach does not respect that number of processed items is always limited and a significant role plays also speed of read/write operations. One general method how to speed up an algorithm is application of space subdivision technique and usually the orthogonal space subdivision is used. In this paper non-orthogonal subdivisions are described. The proposed approach can significantly improve memory consumption and run-time complexity. The proposed modified space subdivision techniques are demonstrated on two simple problems Point-in-Convex Polygon and Point-in-Convex Polyhedron tests.

## I. Introduction

Algorithm's evaluation is quite complicated issue. One approach is based on the Computational Geometry (CG) approach, which deals with the complexity issues expressed by $O(g(N))$ terms, where $g(N)$ is a function, typically. $g(N) = N \lg N$ etc. and $N$ is a number of processed items. It should be noted that $O(N \lg N)$ actually means that the "measured" complexity will be probably something like

$$O(N \lg N) = C_1 N + C_2 \lg N + C_3 \lg^2 N + \cdots + C_k N^2 \lg N$$

where constants $C_1, C_2, \ldots, C_{k-1}$ might be very high. The CG approach evaluates the complexity for $N \to \infty$, which is not a typical engineering problem where algorithms have to optimize for $N \in \langle n_1, n_2 \rangle$. Also the CG approach does not handle problems related to read/write time complexity, caching and speed of the data transfer itself etc. All that mean, that the algorithm with a better computational complexity in the CG sense might be actually slower in real engineering applications for the given interval of $N$, i.e. $N \in \langle n_1, n_2 \rangle$.

Algorithms do have their "optimal" complexity, which is actually the lowest complexity without preprocessing. Of course, for some specific data sets the algorithm might be actually faster.

To speed-up computation a parallel processing can be used. However, it should be noted that parallel processing is actually a "brute force" approach. If we have, e.g. 95% of the code, which can be made in parallel, the final speed-up for an infinite number of processors according to the Amdahl's law will be 20, i.e. the final computation will be 20 times faster but we spend also *infinite resources*!

In the case that many items are processed, the constant part can be preprocessed to speed-up the actual run-time. Space subdivision technique is one of the mostly used approaches beside of the parallel processing.

## II. Space Subdivision

Space subdivision techniques are used in many algorithms across all computational fields. In the following basic types of space subdivision techniques are presented with their fundamental properties.

### A. Orthogonal space subdivision

Orthogonal space subdivision is the simplest and mostly used subdivision techniques. For simplicity, let us consider the $E^2$ case. If we know the Axis Aligned Bounding Box (AABB) of the data set, the AABB is split regularly in one axis, resp. in two axes to smaller cells, where the data, resp. their references are to be stored. No we are getting $M$, resp. $M \times M$, cells, where $N$ items are stored. It means that each item has to be examined and determined a cell, to where the data item is to be stored. So the preprocessing is of the $O(MN)$, resp. $O(M^2 N)$, complexity. In geometric problems an object can interfere with more cells. In the $E^3$ case the preprocessing is $O(M^3 N)$ complexity.

However, there is a possibility how to decrease the memory complexity to $O(N\, dM)$, where $d$ is a dimensionality. However, this technique slightly extend the run-time, details can be found in [4].

In spite of the simplicity, there are some significant disadvantages, especially:

- in the interval of data is unknown, all data have to be read and AABB has to be found
- if large data are to be processed, the preprocessing time is very high due to reading data from an external memory

This work was supported by the Ministry of Education of the Czech Republic, project No.LH12181.

Vaclav Skala is with the University of West Bohemia, Faculty of Applied Sciences, Department .of Computer Science and Engineering, Center of Computer Graphics and Visualization, Univerzitni 8, CZ 30614 Plzen, Czech Republic (http://www.VaclavSkala.eu, http://Graphics.zcu.cz)..

- if only one axis is subdivided, the preprocessing and run-time is quite data sensitive
- it is easy to show that in many cases irregular splitting to cells is needed

However there are some simple problems like

- point-in-convex polygon or point-in-convex polyhedron tests
- convex hull construction in $E^2$, resp. $E^3$

The orthogonal space subdivision causes some problems in preprocessing due to increase of preprocessing complexity with some influences to the run-time computations.

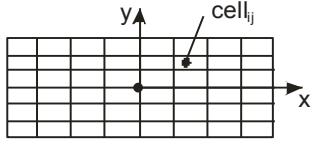

Fig.1 Orthogonal space subdivision

in the case of a convex polygon, the space subdivision is strongly dependent on the polygon rotation, too.

### B. Polar space subdivision

Polar space subdivisions are used in related problems as well. Usually the space is split to regular angular sectors. It means that if the point coordinates are given in the Cartesian coordinates, that an angle and radius has to computed using formulas:

$$r = \sqrt{(x - x_A)^2 + (y - y_A)^2}$$
$$\varphi = \arctan 2(x, y)$$

where $(x_A, y_A)$ is a reference point of the "virtual" origin.

However those formulae do have some weak points, especially:

- radius $r$ has to computed in double precision if coordinates are given is a single precision
- function $\arctan 2(x, y)$ is in *principle imprecise* and giving the relevant angle $\varphi \in \langle 0, 2\pi \rangle$

In the case of the polar space subdivision the computation of $r, \varphi$ is still quite simple in the comparison with the spherical space subdivision.

### C. Spherical space subdivision

In the case of $E^3$ the situations seems to be very similar, however, the formulae are more complex and usually the following formulae are used:

$$r = \sqrt{x^2 + y^2 + z^2}$$
$$\theta = \cos^{-1}\left(\frac{z}{\sqrt{x^2 + y^2 + z^2}}\right)$$
$$\varphi = \tan^{-1}\left(\frac{y}{x}\right)$$

In this case the imprecision of computation is quite high and may lead to wrong selection a cell in the space subdivision.

### D. Cylindrical space subdivision

In the case of the cylindrical space subdivision the situation is simple and is made as a simple modification of the polar space subdivision and subdivision has to be made for the $z$ coordinate, too.

## III. PROPOSED MODIFICATIONS

If the space subdivision is to be effective not only in the preprocessing stage but also during the run-time, all computations have to be as simple as possible. As space subdivision is not actually strictly defined method, it might be imprecise, but reliable for the given purpose giving correct answers all the time.

### A. Modified Polar Space Subdivision

Modification of the original polar space subdivision is quite simple as we do not need strictly regular subdivision, but subdivision close to regular with simple computation. It can be seen that the function $\tan \varphi$ is "nearly" linear for $\varphi \in \langle 0, \pi/4 \rangle$, so the edges of the "virtual" square can be split to segments with the same $\Delta x$, or $\Delta y$ in order to simplify computation of the angular section in which the given point $x$ lies. If $\Delta x = 1$ and $\Delta y = 1$

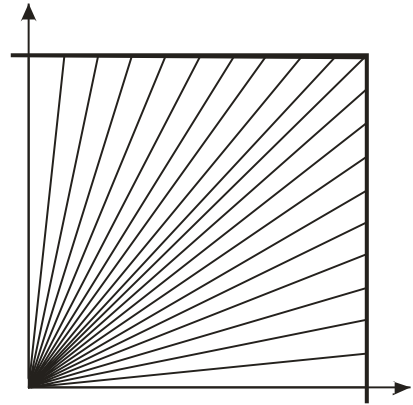

Fig.2 "Virtual" square - 1st quadrant split irregularly

The relevant index $i$ of a segment in the case of the 1st octant for the given point $x = \langle x, y \rangle$ is computed as:

$$i = \frac{y}{x} * m$$

where: $m$ is e number of segments on the "virtual" square edge. Of course, if the point is in the 2nd octant the index has to be computed as:

$$i = \frac{x}{y} * m$$

in order to keep high precision. Other octants are computed similarly.

It can be seen that the computation is much simpler than in the original polar space subdivision approach. Now, there is a question is a similar approach can be applied in the $E^3$ case, i.e. for the spherical space subdivision, as well.

### B. Modified Spherical Space Subdivision

The modification of the polar space subdivision is quite simple as instead of edges split to segments, the faces of a "virtual" cube are split to cells. The advantage of this proposed approach is its simplicity of the index of a cell computation as Δx, Δy, Δz are constant. Indexes of the relevant cells are again quite simple as in principle:

$$i = \frac{y}{x} * m \qquad j = \frac{z}{x} * m$$

and similarly for all faces of the "virtual" cube.

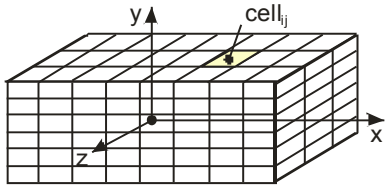

Fig.3 "Virtual" cube splitting

In the $E^3$ case, there is a small complication with indexes as we have to map indexes of cells of the "virtual" cube surface to two dimensional data structure.

### IV. POINT-IN-CONVEX POLYGON WITH $O(1)$ COMPLEXITY

Containment test Point-in-Convex Polygon is a fundamental test in many applications. It is quite a simple test and two algorithms are mostly used. The first one with $O(N)$ complexity is extremely simple to implement. It relies on the edges orientation, e.g. all normal vectors of edges are consistently oriented "out" or "in". This algorithm uses actually a separation by half-planes.

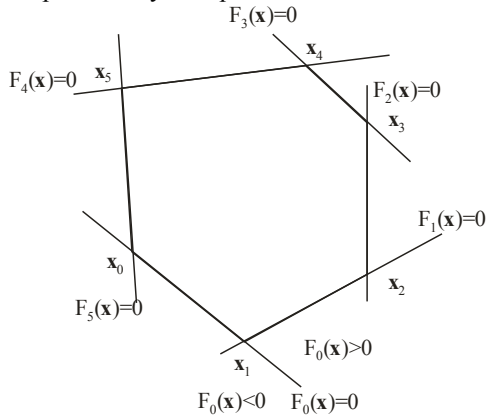

Fig.4 Point-in-Convex Polygon test with $O(N)$ complexity

The second one with $O(\lg N)$ complexity is relatively simple to implement. It relies on the edges orientation as well, but uses vertices ordering, i.e. clockwise or anticlockwise, and binary subdivision on "axis of indexes", i.e. indexes of vertices are binary subdivided, see [12] for details.

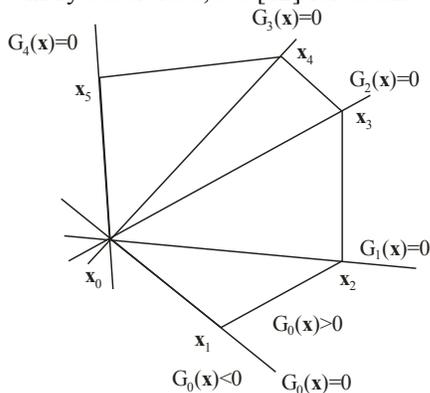

Fig.5 Point-in-Convex Polygon test with $O(\lg N)$ complexity

If the preprocessing can be used, e.g. if many points against a convex constant polygon are to be tested, an algorithm based on one dimensional space subdivision in $y$ axis has been derived recently, Fig.6, and was described in [12]. However this approach has the following problems:

- interval of $y$ axis of all vertices has to be known,
- the memory consumption is highly dependent on geometrical properties, as the width of a slab must be smaller than the smallest edge length in the $y$ coordinate, Fig.7.

It can be seen that the function $\tan \varphi$ is "close" to linear for $\varphi \in \langle 0, \pi/4 \rangle$. In the case of space subdivision we actually do not need *exact regularity*, but "close enough" is acceptable, especially in the case that it leads to some positives.

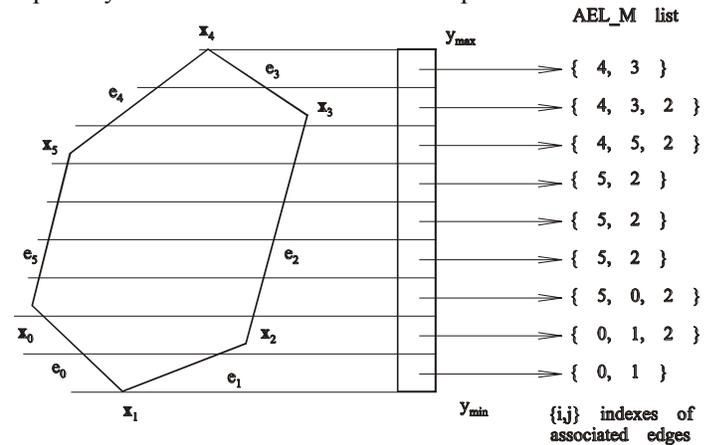

Fig.6 Point-in-Convex Polygon test with $O(1)$ complexity

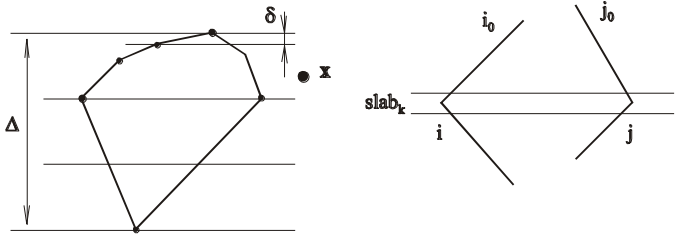

Fig.7 Influence of geometrical properties

Let us consider a convex polygon and let us made a "virtual" square containing the given convex polygon, see Fig.8, for a simplicity (this assumption is not a critical one).

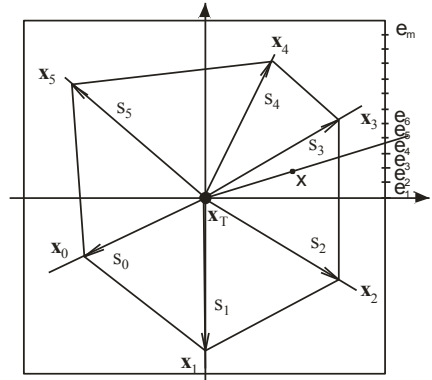

Fig.7 "Virtual" square and the given polygon

Now for the each section on the square edge the relevant edge of the given convex polyhedron can be determined. Due to the limited "resolution", i.e. length of a section, two edges have to be considered if the angular section contains a vertex.

Now, for the given point $x$ the relevant angular section is determined and simple test determine if the point $x$ is inside or outside to the given convex polygon. As the test is based on a half-plane separation test, only 2 addition and 2 multiplication operations are needed.

It can be seen that the test is clearly of $O(1)$ run-time complexity. It can be proved that the section length is determined by the shortest edge length of the given convex polygon.

It should be noted that the proposed algorithm does not need to know AABB box as the "reference" point $x_T$, i.e. the center of the "virtual square" can be computed as:

$$x_T = \left(x_0 + x_{\lfloor N/2 \rfloor}\right)/2$$

The size of the "virtual square" is not actually related to the size of the convex polygon.

## V. Point-in-Convex polyhedron with $O(1)$ complexity

The modification for the $E^3$ case, i.e. for the case of convex polyhedron, the modification is now straightforward. However, there is e a little bit more complex test as we have to test the given point $x$ against planes of the given convex polyhedron incidenting with the relevant spatial angular segment, which is actually of a "pyramid" shape, Fig.3.

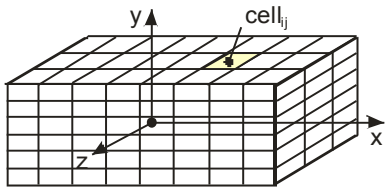

Fig.8 "Virtual" cube for the given polyhedron

Of course, there, there is a legitimate question. How the proposed approach will handle non-self intersecting convex polygon and polyhedron. It seems to that in the non-convex case, the proposed approach is applicable as well, however the expected complexity is of $O(H)$, where H is a number of edges, resp. faces incidenting with the angular segment.

## VI. Conclusion

In this paper new modification of space subdivision techniques based on polar and spherical subdivisions are presented. The proposed approaches are simple to implement and they are robust as well. As a direct consequence algorithms for point-in-convex polygon and point-in-convex polyhedron with $O(1)$ run-time complexity using polar or spherical space subdivision in the preprocessing were briefly described. The algorithms are convenient for application in cases when many points are tested and the given polygon, resp. polyhedron, is constant. It is expected, that the proposed modified polar and spherical space subdivision is widely applicable in many geometry related problems, as well.

## Acknowledgment

The author would like to thank to colleagues at the University of West Bohemia in Plzen and VSB-Technical University, Ostrava for fruitful discussions and to anonymous reviewers for their comments and hints which helped to improve the manuscript significantly.

Vaclav Skala is a Full professor of Computer Science at the University of West Bohemia, Plzen, Czech Republic where he is a full professor of Computer Science. He is the Head of the Center of Computer Graphics and Visualization since 1996.

Vaclav Skala is a member of editorial board of The Visual Computer (Springer), Computers and Graphics (Elsevier), Machine Graphics and Vision (Polish Academy of Sciences), The International Journal of Virtual Reality (IPI Press, USA) and the Editor in Chief of the Journal of WSCG. He has been a member of several international program committees of prestigious conferences and workshops. He is a member of Eurographics Association and he became a Fellow of the Eurographics Association in 2010.

Vaclav Skala has published over 200 research papers in scientific journal and at international research conferences. His current research interests are computer graphics, visualization and mathematics, especially geometrical algebra, algorithms and data structures.
Details can be found at http://www.VaclavSkala.eu